\documentclass[aps,superscriptaddress,twocolumn,amsmath,amssymb]{revtex4}
\usepackage{slashed,amsmath,amssymb,enumitem}
\usepackage[papersize={8.5in,11in}]{geometry}
\usepackage{multirow}
\usepackage{pstricks}
\usepackage{color}
\usepackage{soul}
\usepackage{cancel}
\usepackage{xcolor}

 \normalsize
\def\lsim{\raise0.3ex\hbox{$\;<$\kern-0.75em\raise-1.1ex\hbox{$\sim\;$}}}
\def\gsim{\raise0.3ex\hbox{$\;>$\kern-0.75em\raise-1.1ex\hbox{$\sim\;$}}}

\newcommand{\be}{\begin{eqnarray}}
\newcommand{\ee}{\end{eqnarray}}

\newcommand{\n}{\nonumber\\}

\def\bea{\begin{eqnarray}}
\def\eea{\end{eqnarray}}
\usepackage{graphics}
\usepackage{epsfig}
\usepackage{array}
\usepackage{booktabs}

\begin{document}
\title{ Double Higgs peak in the minimal SUSY $B-L$ model}
\author{W. Abdallah}
\affiliation{Center for Fundamental Physics, Zewail City {of} Science and Technology,
6 October  City, Giza, Egypt.}
\affiliation{Department of Mathematics, Faculty of Science,  Cairo University, Giza, Egypt.}
\author{S. Khalil}
\affiliation{Center for Fundamental Physics, Zewail City {of} Science and Technology,
6 October  City, Giza, Egypt.}
\affiliation{Department of Mathematics, Faculty of Science,  Ain Shams University, Cairo, Egypt.}
\author{S. Moretti}
\affiliation{School of Physics \& Astronomy, University of Southampton, Highfield, Southampton, UK.}
\date{\today}

\begin{abstract}
Motivated by a $\sim 3\sigma$ excess recorded by the CMS experiment at the LHC around a mass of order $\sim 137$ GeV in $ZZ\to 4l$ and $\gamma\gamma$ samples, we analyse the discovery potential of a second neutral Higgs boson in the Supersymmetric $B-L$ extension of the Standard Model (BLSSM) at the CERN machine. We confirm  that a double Higgs peak structure can be generated in this framework, with CP-even Higgs boson masses at  $\sim125$ GeV and $\sim137$ GeV, unlike the case of the Minimal Supersymmetric Standard Model (MSSM).

\end{abstract}
\maketitle
\section{Introduction}
The Higgs boson discovery at the Large Hadron Collider (LHC) in July 2012 has been the beginning of a new era in particle physics. In fact, we now know for certain the mechanism chosen by Nature to generate mass. The actual signals attributed to such a new state, $h$, emerged in a variety of decay modes, $\gamma\gamma$, $ZZ$, $W^+W^-$ as well as $b\bar b$ and
$\tau^+\tau^-$ (in order of decreasing experimental accuracy), all pointing to a Higgs boson mass, $m_h$, of 125 GeV. Furthermore, the 
interaction properties of this new state are rather consistent with those predicted by the Standard Model (SM) and so are
its quantum numbers. 

Intriguingly though, in the search for such a new state, the CMS collaboration also found potential signals for another Higgs boson, $h'$,  with mass $m_{h'}\ge 136.5$ GeV in the ${h'}\to ZZ \to 4l$ decay mode (where $4l$ refers to any possible combination of $e^+e^-$ and $\mu^+\mu^-$ pairs) \cite{Chatrchyan:2013mxa}, wherein a $\sim2\sigma$ excess is appreciable in the vicinities of 145 GeV (specifically, see Figs.~18 and 19 in Ref.~\cite{Chatrchyan:2013mxa}),
and in the ${h'} \to \gamma \gamma$ decay channel, wherein the local $p$-value indicates a possibly significant excess very near
137 GeV at the  $\sim 2.9 \sigma$ level  (see Fig.~6a of the first paper in \cite{CMS:2013wda} and Fig.~2 of the second paper in \cite{CMS:2013wda}). Recall that
these are the two Higgs decays most accurately known by experiment.
We also note, as mentioned in  Ref.~\cite{2higgses}, that various anomalies at $\sim137$ GeV or above have emerged in several other channels, from both ATLAS and CMS at the LHC as well as
CDF and D0 at the Tevatron, see also Ref.~\cite{anomalies}.

An explanation for a second Higgs particle cannot of course be found in the SM. Contrast this with the fact that, in any model of Supersymmetry (SUSY), wherein the SM-like Higgs boson is naturally limited to be at the Electro-Weak (EW) scale (say below $2M_Z$), and where one also finds additional (neutral and CP-even) Higgs bosons \cite{comment}. Thus, there is enough to be tempted to conclude that a SUSY scenario may be behind the aforementioned data. As we shall show, this cannot be its minimal realisation though, the 
so-called Minimal Supersymmetric Standard Model (MSSM). Hence, an explanation for the possibility of a double Higgs peak ought to be found within non-minimal realisations of SUSY.

Amongst the latter, one really ought to single out those that also offer explanations to other data pointing to physics Beyond the SM (BSM), most notably  those indicating that neutrinos oscillate, hence that they have mass. One is therefore well
 motivated in looking at the $B-L$ Supersymmetric Standard Model (BLSSM). The BLSSM is an extension of the MSSM obtained by adopting a $U(1)_{B-L}$ extended  gauge group, i.e., $SU(3)_C \times SU(2)_L \times U(1)_Y \times U(1)_{B-L}$. The particle content of this SUSY $B-L$ model therefore includes the following superfields in addition to those of the MSSM: three SM-singlet chiral superfields, one per generation, $\{\hat{N}_i^c\}_{i=1}^3$ (to be identified with the right-handed neutrinos and their SUSY counterparts), the $\hat{Z}'$ vector superfield necessary to gauge the $U(1)_{B-L}$ symmetry (eventually yielding a physical 
$Z'$ state and its SUSY partner) plus two SM-singlet chiral Higgs superfields $\hat{\chi}_{1,2}$ (finally giving three additional physical Higgs fields and their related SUSY states). In this framework, the scale of $B-L$ symmetry breaking is related to the soft SUSY breaking scale \cite{Khalil:2007dr}. Thus, the right-handed neutrino masses are naturally of order TeV and the Dirac neutrino masses must be less than $10^{-4}$~GeV ({i.e.}, they are of order the electron mass) \cite{Khalil:2006yi1,Khalil:2006yi2,Khalil:2006yi3,Khalil:2006yi4}, as data appear to demand. Furthermore, a similar BLSSM setup, based on an inverse see-saw mechanism, relieves the so-called small hierarchy problem of the MSSM, wherein the discovered Higgs boson mass of 125 GeV is 
dangerously close to its predicted absolute upper limit (130 GeV or so), by providing (s)neutrino mass corrections which
can up-lift this value to 170 GeV or so \cite{BLMSSM-Higgs,sneutrino} (see also \cite{public}). Finally, the BLSSM could also contribute, with respect to the MSSM, additional loop corrections to the $h\to\gamma\gamma$ rate \cite{Basso:2012tr}, should this channel be confirmed to be enhanced by future CERN data.

Notwithstanding this, of particular relevance here is the fact that, in the BLSSM,  
 similarly to the MSSM, the introduction of yet another Higgs state, a singlet $\hat{\chi}_2$ in this case, is necessary in order to cancel the $U(1)_{B-L}$ anomalies produced by the fermionic members of the first Higgs superfield, $\hat{\chi}_1$. 
It will be the mixing of this Higgs singlet with the two Higgs doublets which will relieve the deadlock typical of the MSSM, where a light SM-like Higgs state (at 125 GeV) requires the other Higgs states to be much heavier in comparison, including the CP-even state,   
thereby ultimately being responsible for enabling BLSSM spectra wherein one can find alongside the above SM-like Higgs state
  another rather light physical Higgs boson, $h'$, also CP-even, which can be identified with the object possibly responsible for a second resonance at $\sim137$ GeV.   

In this work, we will set out to prove that this can indeed be the case. The plan of our paper is as follows. In the next section, we describe the Higgs sector of the BLSSM in the presence of a SM-like Higgs state $h$ with a mass of $\sim125$ GeV. In the two following ones, we concentrate in turn on a possible $\sim137$ GeV Higgs signal in $h'\to ZZ\to 4l$ and $h'\to \gamma\gamma$. We conclude in Sect.~4.  

\section{Light Higgs bosons in the BLSSM}
The superpotential of the BLSSM is given by
\bea
\hat{W} &=& Y_u \hat{Q}\hat{H}_2\hat{U}^c + Y_d\hat{Q}\hat{H}_1\hat{D}^c + Y_e\hat{L}\hat{H}_1\hat{E}^c+\mu \hat{H}_1\hat{H}_2\nonumber\\ 
&+&Y_{\nu}\hat{L}\hat{H}_2\hat{N}^c+ Y_N\hat{N}^c\hat{\chi}_1\hat{N}^c +
\mu'\hat{\chi}_1\hat{\chi}_2,
\eea
and, by assuming  universality conditions at the Grand Unification Theory (GUT) scale, we get the SUSY ${B-L}$ soft breaking Lagrangian
\begin{eqnarray}
-\mathcal{L}^{{\text{BLSSM}}}_{\tiny\textnormal{soft}}  &=&   -\mathcal{L}_{\tiny\textnormal{soft}}^{\tiny\textnormal{MSSM}}+  Y_{\nu}^A \tilde{L} H_2 \tilde{N}^c + Y_N^A \tilde{N}^c {\tilde\chi}_1 \tilde{N}^c\n
  &+& m_0^2\left[ |\tilde{E}|^2 + |\tilde{N}|^2 +  |{\tilde\chi}_1|^2 + |{\tilde\chi}_2|^2\right]\n
&+&\left[ B_{\mu'}\,{\tilde\chi}_1 {\tilde\chi}_2  + \frac{1}{2} M_{1/2}\,\tilde{Z}'\tilde{Z}' + h.c.\right],
\end{eqnarray}
where $(Y_f^A)_{ij} = ( Y_f A_0 )_{ij}$ and the tilde denotes the scalar components of the chiral superfields as well as the fermionic components of the vector superfields. We use the same notation for Higgs superfields and their scalar field components. The $U(1)_Y$ and $U(1)_{B-L}$ gauge kinetic mixing can be absorbed in the covariant derivative redefinition. In this basis, one finds
\be
M_Z^2\,\simeq\,\frac{1}{4} (g_1^2 +g_2^2) v^2,  ~~M_{Z'}^2\, \simeq\, g_{BL}^2 v'^2 + \frac{1}{4} \tilde{g}^2 v^2 ,
\ee
where $\tilde{g}$ is the gauge coupling mixing between $U(1)_Y$ and $U(1)_{B-L}$. Furthermore, the mixing angle between $Z$ and $Z'$ is given by 
\be 
\tan 2 \theta'\, \simeq\, \frac{2 \tilde{g}\sqrt{g_1^2+g_2^2}}{\tilde{g}^2 + 16 (\frac{v'}{v})^2 g_{BL}^2 -g_2^2 -g_1^2},
\ee
{where $v=\sqrt{v^2_1+v^2_2}\simeq 246$ GeV and $v'=\sqrt{v'^2_1+v'^2_2}$ with the VEVs of the
Higgs fields given by
\be
\langle{\rm Re} H_i^0\rangle=\frac{v_i}{\sqrt{2}},\;\;\;\;\;\langle{\rm Re} \chi^0_i\rangle=\frac{v'_i}{\sqrt{2}}.
\ee}
The gauge kinetic term induces mixing at tree level between the $H^{{0}}_{1,2}$ and $\chi^{{0}}_{1,2}$ 
states in the BLSSM scalar potential. Therefore, the minimisation conditions of this potential at tree level lead to the following relations \cite{florian2012}: 
\bea
B_\mu &=& -\frac{1}{8} \Big[-2 \tilde{g}g_{BL} v'^2 \cos2\beta' + 4 m_{H_1}^2 - 4 m_{H_2}^2\nonumber\\
      &+& (g_1^2 + \tilde{g}^2 + g_2^2) v^2 \cos 2 \beta \Big] \tan 2 \beta ,\label{Bmu}\\
B_{\mu'} &=& \frac{1}{4} \Big[-2 g^2_{BL} v'^2 \cos2\beta' + 2 m_{\chi_1}^2 - 2 m_{\chi_2}^2\nonumber \\ 
      &+&\tilde{g}g_{BL} v^2 \cos 2 \beta \Big] \tan 2 \beta',
\eea
{where $\tan{\beta}=\frac{v_2}{v_1}$ and $\tan{\beta'}=\frac{v'_1}{v'_2}$.} Note that, with non-vanishing $\tilde{g}$, the $B_\mu$ parameter depends on $v'$ and  the sign of $\cos 2\beta'$. We may have constructive/destructive interference between the first term and other terms in Eq.~(\ref{Bmu}). In general, we find that the typical value of $B_\mu$ is  of order TeV. 

To obtain the masses of the physical neutral Higgs bosons, one makes the usual redefinition of the Higgs fields, i.e.,
$H_{1,2}^0 = {\frac{1}{\sqrt{2}}}(v_{1,2} + \sigma_{1,2} + i \phi_{1,2}) $ and 
$\chi_{1,2}^0 ={\frac{1}{\sqrt{2}}}(v'_{1,2} + \sigma'_{1,2}  + i \phi'_{1,2})$,
where $\sigma_{1,2}= {\rm Re} H_{1,2}^0$, $\phi_{1,2}={\rm Im} H_{1,2}^0$, $\sigma'_{1,2}= {\rm Re} \chi_{1,2}^0$ and $\phi'_{1,2}={\rm Im} \chi_{1,2}^0 $. The real parts correspond to the CP-even Higgs bosons and the imaginary parts correspond to the CP-odd Higgs bosons. The squared-mass matrix of the BLSSM CP-odd neutral Higgs fields at tree level, in the basis $(\phi_1,\phi_2,\phi'_1,\phi'_2)$,  is given by\\
$m^2_{A,A'} = $
\begin{equation}
\left(\begin{array}{cccc}
B_\mu \tan\beta & B_\mu  & 0 & 0 \\
B_\mu  & B_\mu  \cot\beta & 0 & 0 \\
 0 & 0  & B_{\mu'}  \tan\beta' & B_{\mu'}  \\
 0 & 0  & B_{\mu'} & B_{\mu'} \cot\beta' 
\end{array}
\right) \,.
\label{eq:mA2}
\end{equation}
It is clear that the MSSM-like CP-odd Higgs $A$ is decoupled from the BLSSM-like one $A'$ (at tree level). However, due to the dependence of $B_\mu$ on $v'$, one may find $m_A^{{2}} = \frac{2B_\mu}{\sin 2 \beta} \sim m_{A'}^{{2}} =\frac{2 B_{\mu'}}{\sin2 \beta'} \sim {\cal O}(1$ TeV). 

The squared-mass matrix of the BLSSM CP-even neutral Higgs fields at tree level, in the basis $(\sigma_1,\sigma_2,\sigma'_1,\sigma'_2)$,  is given by
\begin{equation}
 M^2 = \left( \begin{array}{cc} 
 			     M^2_{h{H}} ~ &~  M^2_{hh'} \\  \\
                              M^{2^{{T}}}_{hh'} ~ &~  M^2_{h'{H}'} \end{array} \right),
\label{BLMH}
\end{equation}
where $M^2_{h{H}} $ is the usual MSSM neutral CP-even Higgs mass matrix, which leads to the SM-like Higgs boson with mass, at one loop level, of order 125 GeV and a heavy Higgs boson with mass $m_H \sim m_A \sim {\cal O}(1$ TeV). 
In this case, the BLSSM matrix $M^2_{h'{H}'}$  is given by\\ 
$M^2_{h'{H}'}=$
\be
\left( \begin{array}{cc}
               m^2_{A'} c^2_{\beta'} + g^2_{BL} v'^2_1 &-\frac{1}{2} m^2_{A'} s_{2\beta'} - g^2_{BL} v'_1 v'_2 \\
                    \\
              -\frac{1}{2} m^2_{A'}s_{2\beta'} - g^2_{BL} v'_1 v'_2 & m^2_{A'} s^2_{\beta'} + g^2_{BL} v'^2_2
              \end{array}\right),\nonumber
\ee
where $c_x=\cos(x)$ and $s_x=\sin(x)$. Therefore, the eigenvalues of this mass matrix are given by
\bea
{m}^2_{h',H'} &=& \frac{1}{2} \Big[ ( m^2_{A'} + M_{{Z'}}^2 )\n
 &\mp& \sqrt{ ( m^2_{A'} + M_{{Z'}}^2 )^2 - 4 m^2_{A'} M_{{Z'}}^2 \cos^2 2\beta' }\;\Big].\nonumber
\eea
If $\cos^2{{2}\beta'} \ll 1$, one finds that the lightest $B-L$ neutral Higgs is given by %
\be%
{m}_{h'}\; {\simeq}\; \left(\frac{m^2_{A'} M_{{Z'}}^2 \cos^2 2\beta'}{{m^2_{A'}+M_{{Z'}}^2}}\right)^{\frac{1}{2}} \simeq {\cal O}(100~ {\rm GeV}).%
\ee%
The mixing matrix $M_{hh'}^2$ is proportional to $\tilde{g}$ and can be written as \cite{florian2012}
\be
M^2_{hh'}=  \frac{1}{2}\tilde{g} g_{BL} \left( \begin{array}{cc}
              v_1 v'_1 & -  v_1 v'_2\\
                    \\
               ~ - v_2 v'_1 &  ~  v_2 v'_2 
              \end{array}\right).
\ee 
For a gauge coupling $g_{BL} \sim \vert \tilde{g}\vert  \sim {\cal O}(0.5)$, these off-diagonal terms are about one order of magnitude smaller than the diagonal ones. However, they are still crucial for generating interaction vertices between the genuine BLSSM Higgs bosons and the MSSM-like Higgs states. Note that the mixing gauge coupling constant, $\tilde{g}$, is a free parameter that can be positive or negative \cite{florian2012}.


\begin{figure}[t]
\epsfig{file=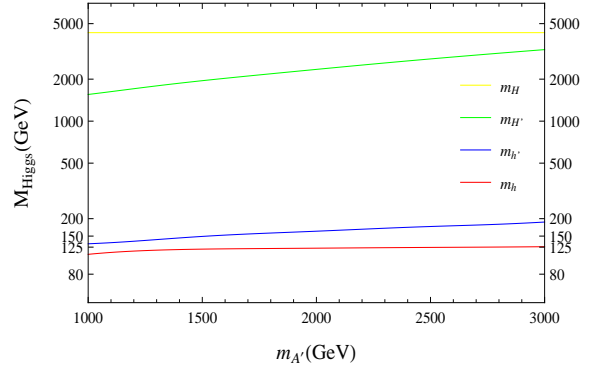,height=5cm,width=8cm,angle=0}
\caption {The BLSSM CP-even Higgs masses versus  $m_{A'}$ for $g_{BL}=0.4$ and $\tilde{g}=-0.4$.} 
\label{fig11}
\end{figure}

In Fig.~\ref{fig11}, we show the masses of the four CP-even Higgs bosons in the BLSSM for $g_{BL}=0.4$ and $\tilde{g}=-0.4$ .  In this plot we fix the lightest MSSM Higgs boson mass to be of order 125 GeV. As can be seen from this figure, as intimated, one of the BLSSM Higgs bosons, $h'$, can be the second lightest Higgs boson ($\sim 137$ GeV).  Both $H$ and $H'$ are quite heavy (since both $m_A$ and $m_{A'}$ are of order TeV). 

Two remarks are in order: firstly, if $\tilde{g}=0$, the coupling of the BLSSM lightest Higgs state, $h'$, with the SM particles will be significantly suppressed ($\leq 10^{-5}$ relative to the SM strength), hence  it cannot account for the considered signals; secondly, in both cases of vanishing and non-vanishing $\tilde{g}$, one may fine-tune the parameters and get a light $m_A$, which leads to a MSSM-like CP-even Higgs state, $H$, with $m_H \sim 137$ GeV. However, it is well known that in the MSSM the coupling $HZZ$ is suppressed with respect to the corresponding one of the SM-like Higgs particle by one order of magnitude due to the smallness of $\cos(\beta -\alpha)$,
where $\sin(\beta -\alpha) \sim 1$. In addition, the total decay width of $H$ is larger than the total decay width of the SM-like Higgs,
$h$, by at least one order of magnitude because it is proportional to $(\cos \alpha/\cos\beta)^2$, which is essentially the square of the coupling of $H$ to the bottom quark. Therefore, the MSSM-like heavy Higgs signal $(pp \to  H \to  ZZ \to 4l)$ has a very suppressed cross section and then it is difficult to probe it at the LHC and it cannot be a candidate for the signals under consideration. 

In the light of this, we will focus in the next section on the lightest BLSSM CP-even Higgs, $h'$, as a possible candidate for the second Higgs peak seen by CMS in Ref.~\cite{CMS:2013wda}. However, before doing so, we ought to setup 
appropriately the BLSSM parameter space, in order to find such a solution.
As mentioned in the introduction, the recent results from CMS  indicate a $\sim 2.9 \sigma$ hint of a second Higgs boson at $137$  GeV or above. Herein, for definiteness, we consider $m_{h'} = 136.5$ GeV as reference BLSSM point. 

\begin{figure}[t]
\epsfig{file=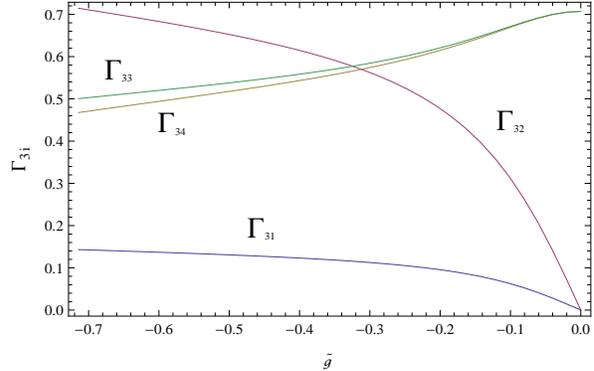,height=5cm,width=8cm,angle=0}
\caption {The mixing of $h'$, $\Gamma_{3i}$, versus the gauge kinetic mixing, $\tilde{g}$.} 
\label{fig5}
\end{figure}

As emphasised above, in the BLSSM, it is quite natural to have two light CP-even Higgs bosons $h$ and $h'$ with mass $125$ GeV and $\sim 137$ GeV, respectively. The CP-even neutral Higgs mass matrix in Eq.~(\ref{BLMH}) can be diagonalised by a unitary transformation:
\be 
{\Gamma}~ M^2 ~ \Gamma^\dag = {\rm diag}\{m_h^2, m_H^2, m_{h'}^2, m_{H'}^2\}. 
\ee
The mixing coupling{s} $\Gamma_{32}$ and $\Gamma_{31}$ are proportional to $\tilde{g}$ and they identically vanish if $\tilde{g}=0$, as one can see in Fig.~\ref{fig5}. Also, in this limit, $\Gamma_{11}$ and $\Gamma_{12}$ approach $\sin \alpha$ and $\cos\alpha$, respectively, where $\alpha$ is the usual CP-even Higgs mixing angle in the MSSM.

The lightest eigenstate $h$ is the SM-like Higgs boson, for which we will fix its mass to be exactly 125 GeV. As mentioned, the numerical scans confirm that the $h'$ state can then be the second light Higgs boson with mass of ${\cal O}$(137 GeV). The other two CP-even states, $H, H'$, are heavy (of ${\cal O}(1)$ TeV). The $h'$ can be written in terms of  gauge eigenstates as 
\be 
h' = \Gamma_{31} ~\sigma_1 + \Gamma_{32} ~\sigma_2 + \Gamma_{33}~ \sigma'_1  +\Gamma_{34}~ \sigma'_2.
\ee 
Thus, the couplings of the $h'$ with up- and down-quarks  are given by 
\be%
h'\, u\, \bar{u} :~ {-i}~\frac{m_u}{v} \frac{\Gamma_{{32}}}{\sin \beta}, ~~~~~  h'\, d\, \bar{d} : ~ {-i}~\frac{m_d}{v} \frac{\Gamma_{{31}}}{\cos \beta}.
\ee
Similarly, one can derive the $h'$ couplings with the ${W^{+}}{W^-}$ and $Z{Z}$ gauge bosons:
\bea%
h'\,W^+\,W^- &:& {i}~g_{{2}} M_W \left(\Gamma_{{32}} \sin \beta + \Gamma_{{31}} \cos\beta\right),\n
h' Z Z &:&\frac{i}{2}\Big[4 g_{BL}\sin^2{\theta'}\left(v'_1\Gamma_{32} + v'_2\Gamma_{31}\right)\n
&+&\left(v_2\Gamma_{32} + v_1\Gamma_{31}\right)\left(g_z \cos{\theta'}-\tilde{g}\sin{\theta'}\right)^2\Big].\nonumber
\eea
{Since $\sin{\theta'}\ll 1$, the coupling of the $h'$ with $ZZ$, $g_{h'ZZ}$, will be as follows}
\be
g_{h'ZZ}\simeq i~g_z\, M_Z  \left(\Gamma_{32} \sin{\beta}+ \Gamma_{31} \cos{\beta}\right),
\ee
where $g_z=\sqrt{g_1^2+g_2^2}$. In our analysis we have used SARAH \cite{florianSARAH} and SPheno \cite{PorodSPheno,florianSPheno} to build the BLSSM. Furthermore, the matrix-element calculation and event generation were derived from MadGraph 5 \cite{Madgraph5}
and manipulated with MadAnalysis 5 \cite{Madanalysis5}. Finally, notice that all current experimental constraints, from both collider (LEP2, Tevatron and LHC) and flavour  (LHCb, BaBar and Belle) are taken into account in our numerical
scans.

In what follows, we will consider the BLSSM benchmark point for soft SUSY breaking parameters given in Tab.~\ref{tab1}.

\begin{table}[t]
{\small\fontsize{7}{7}\selectfont{
\begin{tabular}{|c|c|c|c|c|c|}
\hline
\multicolumn{6}{|c|}{Inputs} \\
\hline
$g_{BL}$ & $\tilde{g}$ & $\tan{\beta}$ & $\tan{\beta'}$ & $M_{Z'}$ &  $m^2_{H_1}$\\
\hline
$0.55$ & $-0.12$ & $5$ & $1.15$ & $1700$ & $1.1\times 10^6$ \\
\hline
$m^2_{H_2}$ & $m^2_{\chi_1}$& $m^2_{\chi_2}$& $Y_\nu^{\text{diag.}}$ & $Y_N^{\text{diag.}}$& \text{sign}$(\mu,\mu')$\\
\hline
$-1\times 10^7$ & $-2.8\times 10^4$ & $7.8 \times 10^5$ & $10^{-4}$ & $0.43$& 1\\
\hline
$(m^2_{\tilde{q}})^{\text{diag.}}$ & $( m^2_{\tilde{\ell}})^{\text{diag.}}$ & $(m^2_{\tilde{d}})^{\text{diag.}}$& $(m^2_{\tilde{u}})^{\text{diag.}}$ & $(m^2_{\tilde{e}}    )^{\text{diag.}}$ & $(m^2_\nu)^{\text{diag.}}$\\
\hline
$3.9\times 10^7$ & $3.1\times 10^5$ & $4\times 10^7$ & $4\times 10^7$ & $1.8\times 10^5$& $7.9\times 10^5$\\
\hline
\multicolumn{6}{|c|}{Outputs}\\
\hline
 $m_h$& $m_{h'}$& $m_H$& $m_{H'}$& $m_A$& $m_{A'}$ \\
 \hline
125  & 136.5 & $3.1\times 10^3$  & $2.3\times 10^3$  & $3.1\times 10^3$  & $1.6\times 10^3$  \\
\hline
\end{tabular}}}
\caption{BLSSM benchmark point in terms of inputs (to SARAH and SPheno) and outputs (used in our analysis). (Dimensions of
masses (squared) are GeV (GeV$^2$).)}
\label{tab1} 
\end{table} 
%
\section{Search for a $\sim 137$ GeV Higgs boson in ${h'}\to ZZ \to 4l$}
%
The Higgs decay into $ZZ \to 4l$ is one of the golden channels, with low background, to search for Higgs boson(s). The search is performed by looking for resonant peaks in the $\text{m}_{4l}$  spectrum, i.e.,
the invariant mass of the $4l$ system. In CMS \cite{Chatrchyan:2013mxa}, this decay channel shows two significant peaks at 125 GeV and above $137$ GeV. We define by $\sigma(pp \to h')$  the total $h'$ production cross section, dominated by gluon-gluon fusion, and by $\text{BR}(h'\to ZZ)$ the $h'$ Branching Ratio (BR) in the $ZZ$ channel. From the previous section, it is then clear that 
\be
\frac{\sigma(pp \to h')}{\sigma(pp \to h)^{\text{SM}}} \simeq \left ( \frac{\Gamma_{{32}}}{\sin \beta}\right )^2,
\ee
(wherein the label SM identifies the SM Higgs rates computed for a 125 GeV mass),
which, for $m_{h'} \approx137$ GeV, is of order {${\cal O}(0.1)$}. Also the ratio between BRs can be estimated as 
$$\frac{\text{BR}( h'\to ZZ)}{\text{BR}(h\to ZZ)^{\text{SM}}} \simeq \left(1+\frac{\Gamma^{\text{SM}}_{h\rightarrow W W^*}}{\Gamma^{\text{SM}}_{h\rightarrow b \bar{b}}}\right)\frac{F(M_Z/m_{h'})}{F(M_Z/m_{h})^{\rm SM}}\nonumber
$$
 $$\times\left[\left(\frac{\Gamma_{31}\sec{\beta}}{\Gamma_{32} \sin{\beta}+ \Gamma_{31} \cos{\beta}}\right)^2+2F\left(\frac{M_W}{m_{h'}}\right)\right]^{-1},$$
where 
$$ F(x)=\frac{3(1-8 x^2+20 x^4)}{(4 x^2-1)^{1/2}}\arccos\left(\frac{3 x^2-1}{2 x^{3}}\right)$$
 $$-\frac{1-x^2}{2 x^2}(2-13 x^2+47 x^4)-\frac{3}{2}(1-6 x^2 +4 x^4)\log{x^2}.$$
\begin{figure}[t]
\epsfig{file=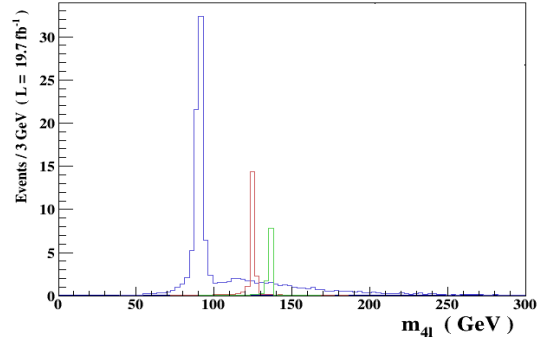,height=5cm,width=8cm,angle=0}
\caption {The number of events of the processes $pp \to Z\to 2l \gamma^* \to 4l$ (blue), $pp \to h \to ZZ \to 4l$ (red) and $pp \to h' \to ZZ \to 4l$ (green) versus the invariant mass of the out going particles (4-leptons), $\text{m}_{4l}$.} 
\label{fig2}
\end{figure}
Now we analyse the kinematic search for the BLSSM Higgs boson, $h'$, in the decay channel to $ ZZ \to 4l$.  
In Fig.~\ref{fig2}, we show the invariant mass of the 4-lepton final state from $ pp \to h' \to ZZ \to 4l$ at $\sqrt{s}={8}$ TeV, after applying a $p_T$ cut of 5 GeV on the four leptons. The SM model backgrounds from the $Z$ and {125} GeV Higgs boson decays,  $pp \to Z\to 2l \gamma^* \to 4 l$ and $pp \to h \to ZZ \to 4l$, respectively, are taken into account, as
demonstrated by the first two peaks in the plot (with the same $p_T$ requirement). It is clear that the third peak at $\text{m}_{4l} \sim 137$ GeV, produced by the decay of the BLSSM Higgs boson $h'$ into $ZZ \to 4 l$, can reasonably well account for the events observed by CMS \cite{Chatrchyan:2013mxa} with the 8 TeV data. This is shown in Tab.~\ref{tab2}, where the mass interval in m$_{4l}$ that we have investigated to extract the $h'$ signal
is wide enough to capture the prominent 145 GeV anomaly seen in CMS. 
\begin{table}[t]
\begin{tabular}{|c|c|c|c|c|}
\hline
\multicolumn{5}{|c|}{ Number of events for $ 19.7\; \text{fb}^{-1}$ at $\sqrt{s}=8$ TeV}\\
\hline
\multirow{2}{*}{Higgs mass}&Observed&Expected&\multicolumn{2}{c|}{Background}\\
\cline{4-5}
& (CMS)&(BLSSM) &$Z\to 2l\gamma^*$&$h \to ZZ$\\
\hline
125 GeV & 25 & 18.5&6.6&-\\
\hline
136.5 GeV & 29 & 10.2&9.15&0.8\\
\hline
\end{tabular}
\caption{The observed (by CMS) and expected (from the BLSSM) number of events in a mass window around $m_h=$ 125 GeV ($121$ GeV $<m_{4l}< 131 $ GeV) and $m_{h'}=136.5$ GeV ($ 131$ GeV $<m_{4l}< 152 $ GeV) in the $ZZ\to 4l$ channel compared to the expected (dominant) $pp\to Z \to 2l\gamma^*\to 4l$ and $pp \to h \to ZZ \to 4l$ backgrounds.}
\label{tab2} 	
\end{table} 
\section{Search for a $\sim 137$ GeV Higgs boson in $h'\to \gamma \gamma$}
Now we turn to the di-photon channel, which provides the greatest sensitivity for Higgs boson discovery in the
intermediate mass range (i.e., for Higgs masses below $2M_W$). Like the SM-like Higgs, the $h'$ decays into two photons through a triangle-loop diagram dominated by (primarily) $W$ and (in part) top quark exchanges.  As shown in a previous section, the couplings of the $h'$ with top quarks and $W$ gauge bosons are proportional to some combinations of $\Gamma_{{31}}$ and $\Gamma_{{32}}$, which may then lead to some suppression in the partial width $\Gamma(h' \to  \gamma \gamma)$. In the SM,  $\text{BR}(h\to \gamma \gamma) \simeq 2\times 10^{-3}$. Similarly, in the BLSSM, we found that, for $m_{h'}= 136.5$ GeV, the BR of $h'$ in photons amounts to {$2.15 \times 10^{-3}$}. 
 \begin{figure}[t]
\epsfig{file=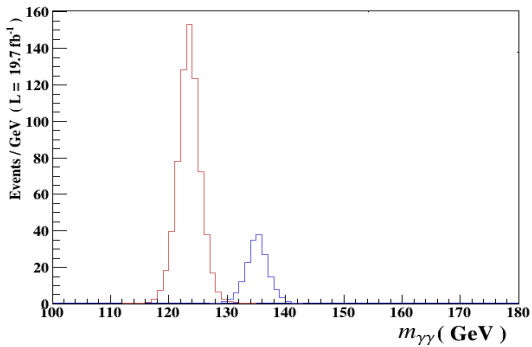,height=5cm,width=8cm,angle=0}
\caption { The number of events of the processes $pp \to h \to \gamma \gamma$ (red), $pp \to h' \to \gamma \gamma$ (blue) versus the invariant mass of the outgoing particles (di-photons), $m_{\gamma\gamma}$.} 
\label{fig3}
\end{figure}
\begin{table}[t]
\begin{tabular}{|c|c|c|}
\hline
\multicolumn{3}{|c|}{Number of events for $19.7\; \text{fb}^{-1}$ at $\sqrt{s}=8$ TeV}\\
\hline
{Higgs mass}& Observed (CMS) & Expected (BLSSM)\\
\hline
$125$ GeV & 610 & 666\\
\hline
$136.5$ GeV & 170 &177 \\
\hline
\end{tabular}
\caption{The observed (by CMS) and expected (from the BLSSM) number of events (after subtracting background) in a mass window around $m_h=$ 125 GeV ($120$ GeV $<{m_{\gamma\gamma}}< 130 $ GeV) and $m_{h'}=136.5$ GeV ($ 131$ GeV $<{m_{\gamma\gamma}}< 141 $ GeV) in the $\gamma\gamma$ channel.}
\label{tab3} 
\end{table} 
The distribution of the di-photon invariant mass is presented in Fig.~\ref{fig3} for a centre-of-mass energy $\sqrt{s}={8}$ TeV. Again, here, the observed $h\to \gamma\gamma$ SM-like signal around 125 GeV is taken as background while the $Z\to \gamma\gamma$ background can now be ignored \cite{ZAA}. As expected, the sensitivity to the $h'$ Higgs boson is severely reduced with respect to the presence of the already observed Higgs boson, yet a peak is clearly seen at 136.5 GeV and is very compatible with the excess seen by CMS \cite{CMS:2013wda}. This is shown in Tab.~\ref{tab3}. It is worth mentioning that here we consider both the gluon-gluon fusion and the vector-boson fusion modes for both $h$ and $h'$ production. Before closing, we should also mention that the $h'\to\gamma\gamma$ enhancement found in the BLSSM may be mirrored in the $\gamma Z$ decay channel for which, at present, there exists some constraints, albeit not as severe as
in the $\gamma\gamma$ case. We can anticipate  \cite{progress} that the BLSSM regions of parameter space studied
here are consistent with all available data \cite{HZgamma}.
\vspace*{1.0truecm}
\section{Conclusions}
In summary, if the $\sim2.9\sigma$ hint of a second Higgs peak at $\sim137$ GeV or above seen by CMS in both 
$ZZ\to4l$ and $\gamma\gamma$ (and also hinted by other measurements from other experiments)
after the 7 and 8 TeV runs of the LHC were to be confirmed at 13 and/or 14 TeV,
this would rule out the MSSM as an explanation while electing the BLSSM to be a possible candidate. We have illustrated this
by using a benchmark point over the BLSSM parameter space, though we have verified that this is naturally possible over
large expanses of it, as we have scanned over the relevant BLSSM parameters, all such situations being compliant with current experimental bounds.
So far ATLAS have made no conclusive statement about the existence of such a possible second Higgs peak, so that their 
forthcoming data will be even more crucial in order to make a final assessment  about the scenario we have investigated.   
\section*{Acknowledgements}
The work of W.A. and S.K. is partially supported by ICTP grant AC-80. S.M. is partially supported through the NExT Institute. W.A. and S.K. would like to thank N. De Filippis, A. Ali and R. Aly for fruitful discussions.


\begin{thebibliography}{99}
\bibitem{comment}
As we are intending to explain the aforementioned $ZZ$ excess at $\sim 137$ GeV via a $h'$ state, the latter can only be CP-even.
\end{thebibliography}
\end{document}